\documentclass[fleqn,10pt]{wlscirep}
\usepackage[utf8]{inputenc}
\usepackage[T1]{fontenc}
\title{Comparative analysis of privacy-preserving open-source LLMs regarding extraction of diagnostic information from clinical CMR imaging reports}

\author[1, 2, *]{Sina Amirrajab PhD}
\author[2]{Volker Vehof MD}
\author[2]{Michael Bietenbeck PhD}
\author[2]{Ali Yilmaz MD}
\affil[1]{The D-Lab, Department of Precision Medicine, GROW - Research Institute for Oncology and Reproduction, Maastricht University, Maastricht, the Netherlands. }
\affil[2]{Division of Cardiovascular Imaging, Department of Cardiology I, University Hospital Münster, Von-Esmarch-Str. 48, 48149 Münster, Germany}

\begin{abstract}

We investigated the utilization of privacy-preserving, locally-deployed, open-source Large Language Models (LLMs) to extract diagnostic information from free-text cardiovascular magnetic resonance (CMR) reports. We evaluated nine open-source LLMs on their ability to identify diagnoses and classify patients into various cardiac diagnostic categories based on descriptive findings in 109 clinical CMR reports. Performance was quantified using standard classification metrics including accuracy, precision, recall, and F1 score. We also employed confusion matrices to examine patterns of misclassification across models. Most open-source LLMs demonstrated exceptional performance in classifying reports into different diagnostic categories. Google's Gemma 2 model achieved the highest average F1 score of 0.98, followed by Qwen2.5:32B and DeepseekR1-32B with F1 scores of 0.96 and 0.95, respectively. All other evaluated models attained average scores above 0.93, with Mistral and DeepseekR1-7B being the only exceptions. The top four LLMs outperformed our board-certified cardiologist (F1 score of 0.94) across all evaluation metrics in analyzing CMR reports. Our findings demonstrate the feasibility of implementing open-source, privacy-preserving LLMs in clinical settings for automated analysis of imaging reports, enabling accurate, fast and resource-efficient diagnostic categorization.

\end{abstract}
\begin{document}

\flushbottom
\maketitle
%
%
\thispagestyle{empty}


\section*{Introduction}

Cardiovascular magnetic resonance (CMR) imaging has emerged as a cornerstone of non-invasive cardiovascular diagnostics, offering unparalleled advantages for assessing cardiac structure, function, and tissue characteristics. Unlike other imaging modalities, CMR provides unique insights into myocardial viability, inflammation, infiltration and fibrosis \cite{mcdonagh2021esc, ponikowski2016esc}. 

CMR interpretation requires synthesizing complex imaging data into focused and clinically meaningful reports. These reports typically include structured sections on patient history, imaging protocols, quantitative measurements, qualitative observations and – if possible – a final (suspected) diagnosis \cite{hundley2022scmr}. While structured reporting templates promoted by the Society for Cardiovascular Magnetic Resonance (SCMR) improve consistency, many clinicians still rely on narrative to capture nuanced observations and to describe possible differential diagnoses, leading to heterogeneity in format, terminology, and detail \cite{bunck2020structured}. This variability complicates data curation for research, quality assurance, and longitudinal patient monitoring.

Large Language Models (LLMs) are rapidly transforming numerous fields, and healthcare is no exception, with their capacity to revolutionize medical report analysis \cite{zhang2025revolutionizing}. Recent research underscores the potential of LLMs in diverse medical applications, from clinical note summarization to assisting in diagnoses by analyzing complex patient data.  LLMs like ChatGPT are being leveraged in clinical settings, demonstrating both impressive capabilities and areas for further development in their integration into healthcare \cite{thirunavukarasu2023large, meng2024application, clusmann2023future, park2024assessing}. 

The application of LLMs to “clinical report analysis” is associated with substantial challenges, particularly concerning patient privacy and data security. Closed-source LLMs, such as OpenAI's ChatGPT, often necessitate accessing their capabilities through APIs, which inherently involves sharing patient data with external servers. This data transfer raises substantial compliance issues with General Data Protection Regulation (GDPR), as sensitive patient information may be processed outside of secure, local environments.

An alternative approach to mitigate these privacy concerns involves the deployment of locally run, open-source LLMs within hospital infrastructures. This strategy offers several advantages over cloud-based models. Primarily, processing clinical reports with a local LLM keeps patient data within the hospital's secure network, significantly reducing the risk of external data breaches and facilitating compliance with stringent regulations like GDPR \cite{gilbert2023large, pauws2025bridging}.

The present feasibility study aims to demonstrate the use of open-source LLMs in clinical settings for secure and efficient analysis of text-based CMR imaging reports. By evaluating the performance of locally deployed LLMs on these reports, we showcase an alternative to cloud-based solutions, which often raise privacy concerns when handling sensitive patient data.

\section*{Results}

\begin{figure}[ht]
\centering
\includegraphics[page=3, width=\linewidth]{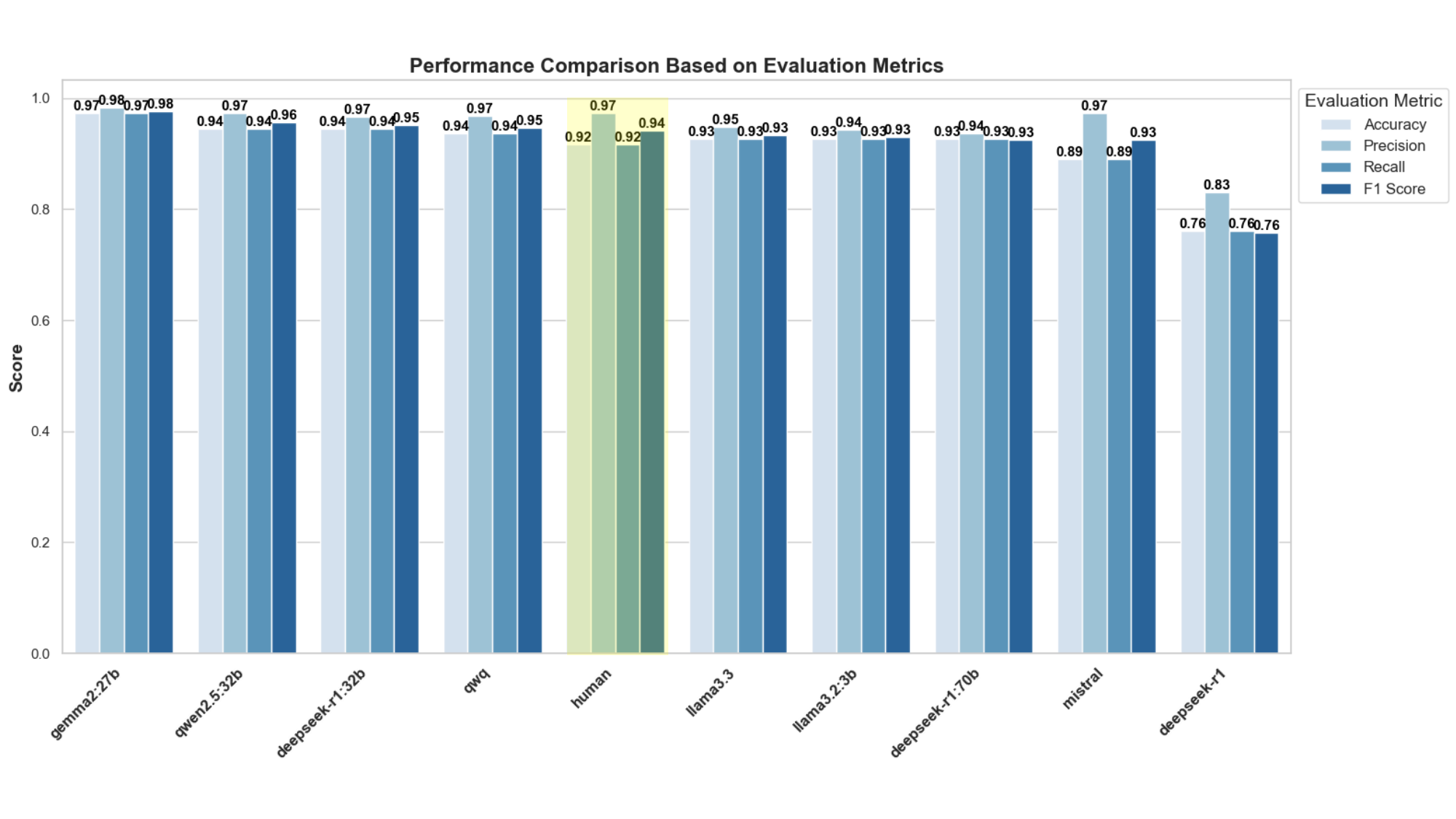} 
\caption{LLM comparison in terms of different evaluation metrics. The best model is determined based on the highest mean score across metrics ordered from left to right. The performance of human evaluator is highlighted. }
\label{fig:;performance}
\end{figure}

A detailed analysis of model performance in CMR imaging report classification revealed consistently strong results: The bar chart in Figure \ref{fig:;performance}  compares nine LLMs, showing that most achieved excellent accuracy, precision, recall, and F1 scores above 0.93. The Gemma-2-27b model demonstrated the highest overall performance, with an average score exceeding 0.97 across all metrics. It was closely followed by Qwen2.5:32b and DeepSeek-R1:32b, which attained F1 scores of 0.96 and 0.95, respectively. The reasoning model of Qwen2.5, the QWQ model, achieved a comparable score to, and slightly higher than, that of our cardiologist. 

\begin{figure}[ht]
\centering
\includegraphics[page=4, width=\linewidth]{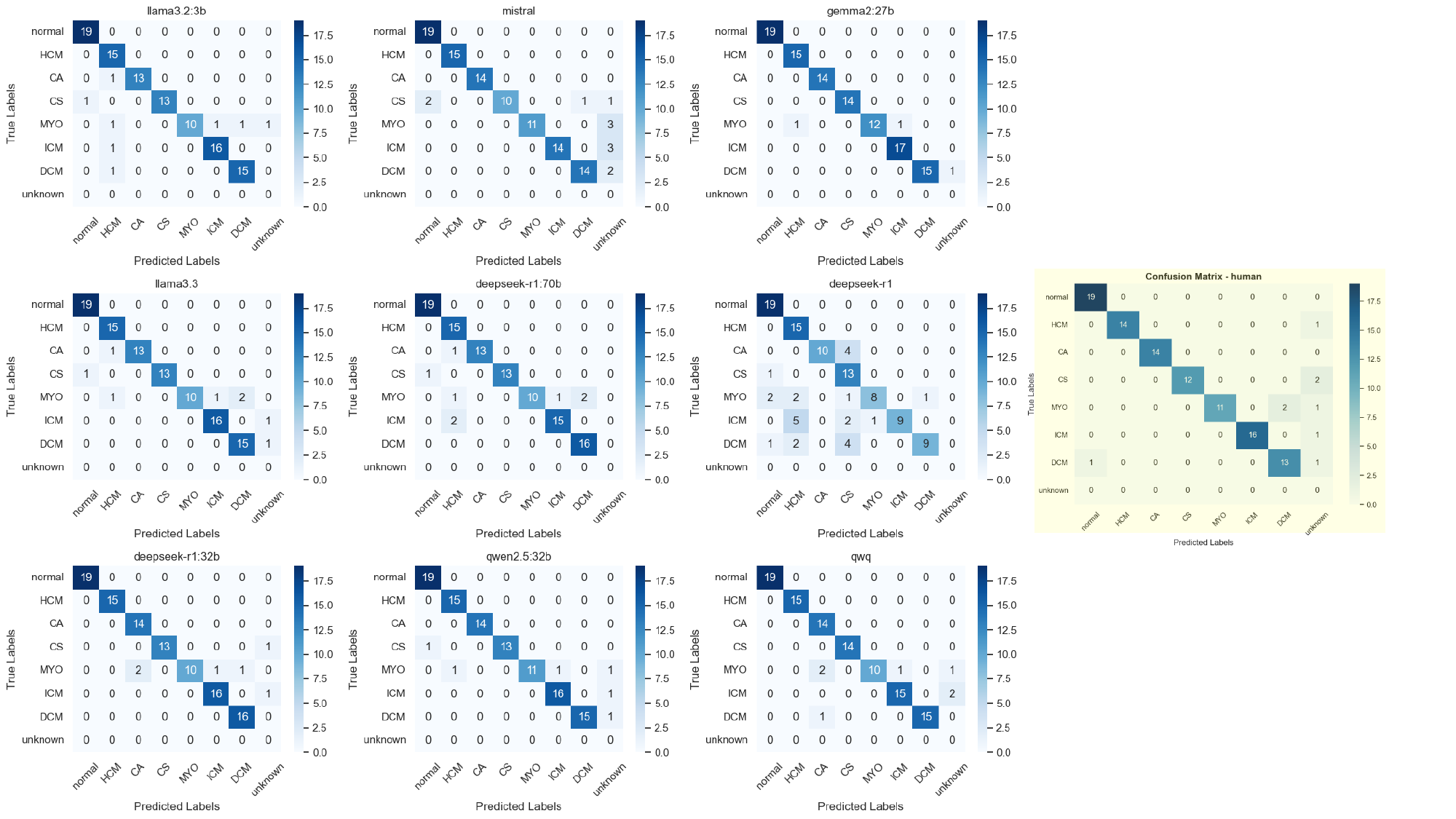} 
\caption{depicts confusion matrices illustrating model performance and human performance in classifying eight cardiac diagnostic categories from unstructured imaging reports. Strong diagonal dominance in Gemma2-27b, DeepSeek-R1-32b, and Qwen2-5-32b, indicates high accuracy, with occasional misclassifications observed for myocarditis (MYO).}
\label{fig:;confusion}
\end{figure}

The Mistral model performed slightly lower, with scores around 0.89, while DeepSeek-V1 had the weakest performance at approximately 0.76. Most models exhibited consistent performance across all four metrics, indicating balanced classification capabilities with minimal trade-offs. These findings suggest that multiple LLMs can effectively extract diagnostic categories from text-based cardiac imaging reports, offering potential applications in automating imaging report processing and supporting clinical decision-making in cardiology.
Figure \ref{fig:;confusion} presents confusion matrices that visualize each model's classification performance across eight cardiac diagnostic categories based on text-based CMR imaging reports. These matrices illustrate the relationship between true diagnostic labels (rows) and predicted labels (columns), with diagonal cells representing correct classifications and off-diagonal cells indicating errors. The intensity of the blue coloration reflects the number of cases in each cell.

Overall, most models showed a strong diagonal dominance, particularly Gemma2-27b, DeepSeek-R1-32b, Qwen2-5-32b, and LLaMA3.3, indicating high classification accuracy. Normal cardiac findings and HCM were consistently classified with perfect or near-perfect accuracy. However, some categories showed characteristic misclassification patterns: For example, MYO was occasionally misclassified as ischemic (ICM) or dilated cardiomyopathy (DCM), while CS was sometimes confused with other conditions, particularly in the mistral model.
The DeepSeek-R1 model showed the highest rate of misclassification, especially for cardiac amyloidosis (CA), myocarditis, ICM and DCM. These patterns underscore the strengths of top-performing models in distinguishing cardiac conditions from unstructured or less structured (narrative) text while highlighting persistent challenges in differentiating pathologically similar diseases. Myocarditis, in particular, was frequently misclassified as other cardiomyopathies, suggesting inherent difficulties in its classification, even for our human evaluator.

\begin{figure}[ht]
\centering
\includegraphics[page=5, width=0.9\linewidth]{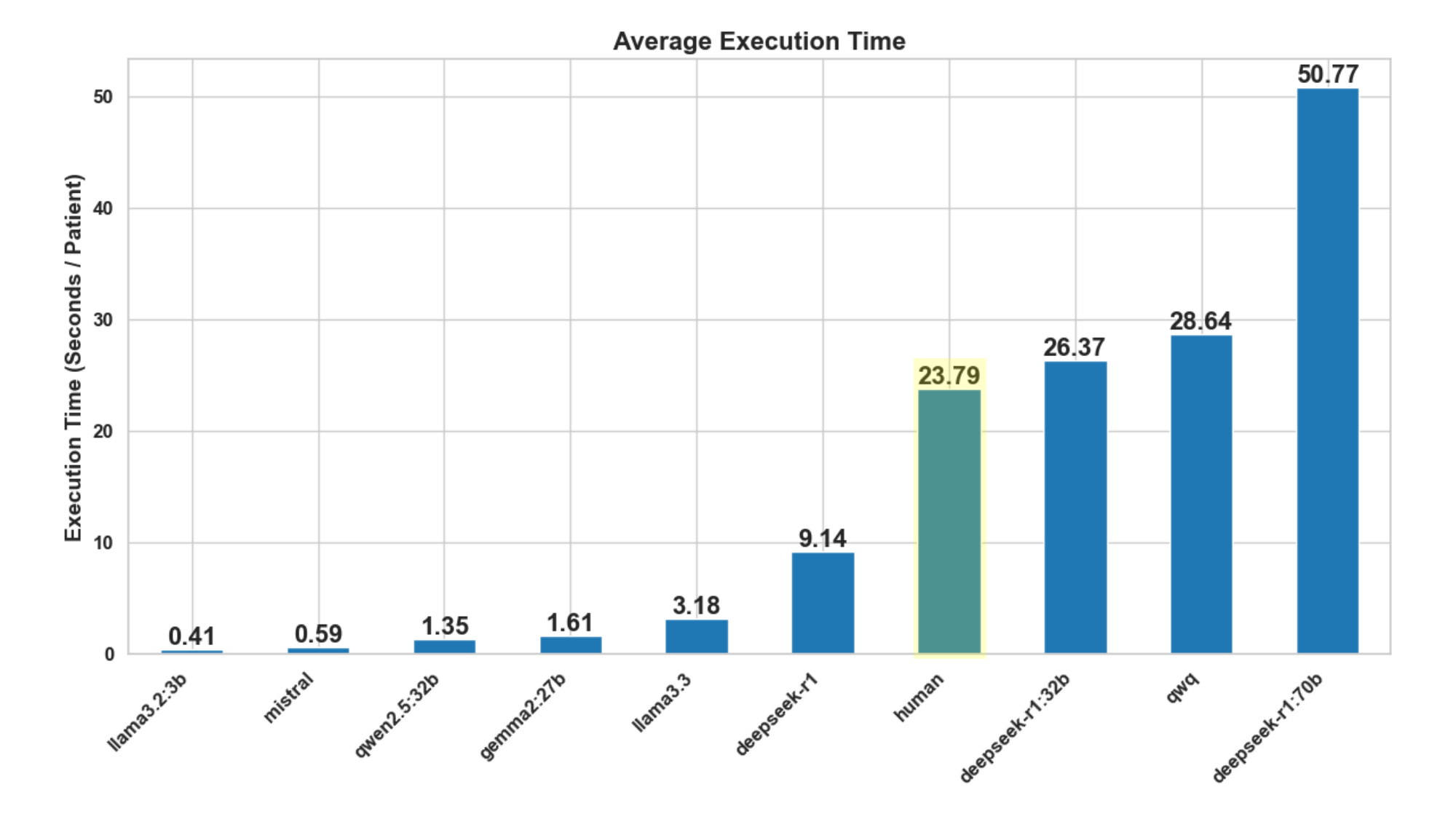} 
\caption{shows the execution time (seconds per patient) for each LLM and human evaluator, highlighting significant efficiency differences. Smaller models like LLaMA 3 (2.3B) and Mistral were the fastest, while larger models, such as DeepSeek-R1 (70B), had the highest execution time without a clear accuracy advantage.}
\label{fig:;time}
\end{figure}

Figure \ref{fig:;time} shows the execution time (seconds per patient) for each LLM evaluated and average time for our cardiologist evaluator. The results reveal significant differences in computational efficiency. The fastest model, LLaMA 3 (2.3B), had an average execution time of 0.41 seconds per patient, followed by Mistral (0.59s) and Qwen 2.5 (32B) (1.35s). These models exhibit high computational efficiency, making them well-suited for real-time or high-throughput clinical applications. Gemma 2 (2.7B) (1.61s) and LLaMA 3 (3B) (3.18s) also demonstrated relatively low execution times, suggesting that smaller or optimized architectures can process reports with minimal latency.

\section*{Discussion}

This study demonstrates the effectiveness of open-source, locally run LLMs in analyzing CMR imaging reports to automatically and quickly identify patient diagnoses based on descriptive report findings. Our present results indicate that most LLMs achieve exceptionally high scores across various evaluation metrics, highlighting their potential for automated diagnosis extraction from text-based imaging reports, circumventing the need for labor-intensive supervised training data curation. 

The top four LLMs outperformed our board-certified cardiologist across all evaluation metrics in analyzing and classifying CMR reports. Notably, larger models like DeepSeek-R1 (32B) and qwq required significantly higher execution times (26.37s and 28.64s), with DeepSeek-R1 (70B) recording the highest at 50.77 seconds per patient. Our evaluator spent a similar time of 23.79 seconds per patient. 

Importantly, DeepSeek-R1 (70B) that had the highest execution time, performed worse in classification accuracy, indicating that increased computation does not necessarily improve performance. These results underscore the need to balance model complexity with processing efficiency, particularly in clinical settings where timely decision-making is essential.

One limitation of this study is the limited number of imaging reports used to evaluate LLM performance, as well as the reliance on single-center data. As a next step, we aim to analyze more than 4,000 cases using the three best-performing LLMs to generate a consensus result.  

This study demonstrates the feasibility of using open-source, locally run LLMs to analyze CMR reports and derive diagnostic conclusions from text-based clinical data while ensuring privacy. Our findings show that existing LLMs enable classification performances above 0.93 across most evaluation metrics, highlighting their effectiveness and usefulness in automating the analysis of imaging reports.

\section*{Methods}
\begin{figure}[ht]
\centering
\includegraphics[page=2, width=\linewidth]{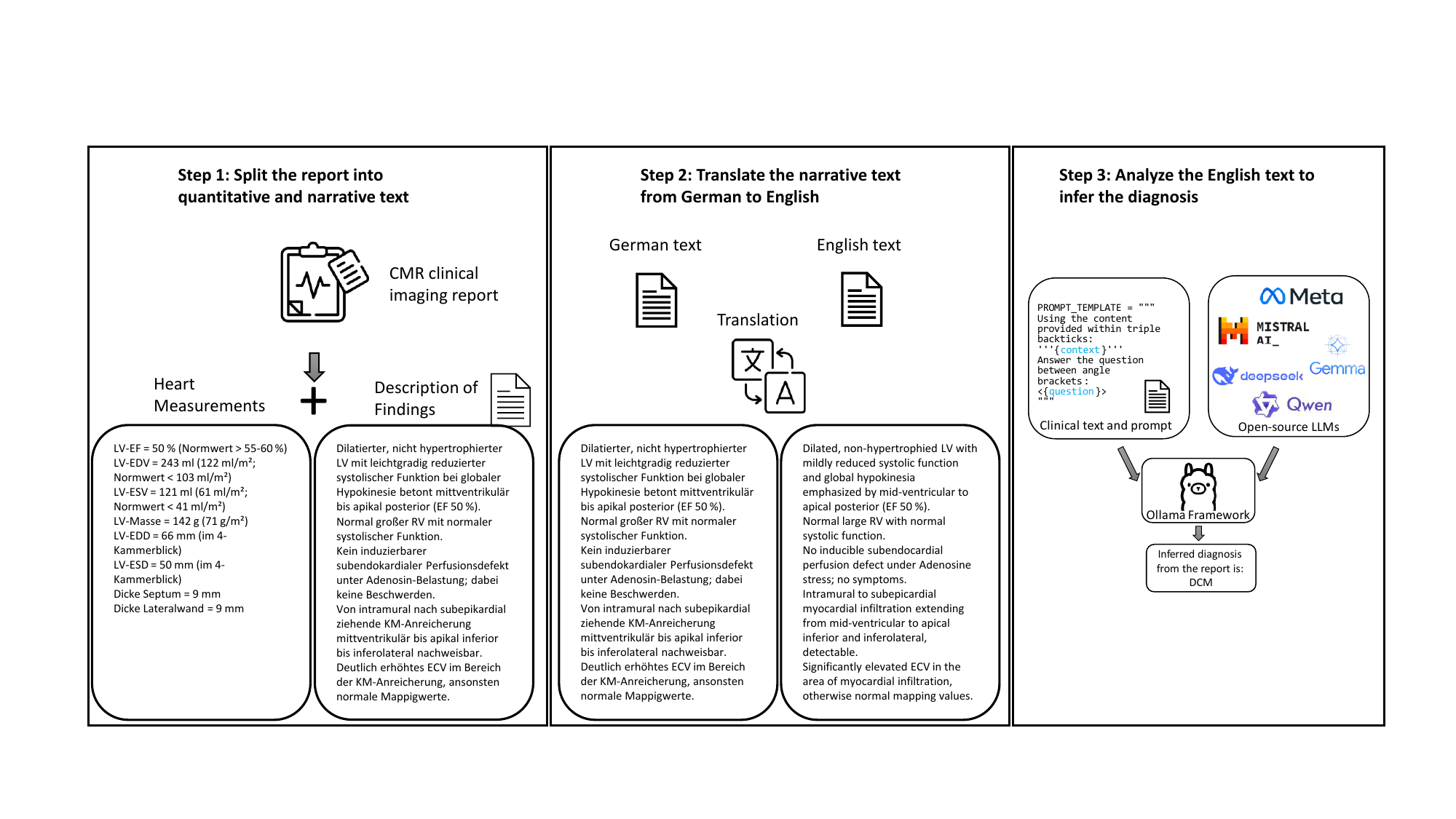} 
\caption{The proposed pipeline for analyzing imaging reports to automatically extract diagnostic information from the text, including three steps to 1) split clinical reports into a first quantitative and a second narrative information, 2) translate the narrative text from German to English, and 3) analyze the English text with open-source locally-run LLMs to infer the diagnosis.}
\label{fig:;method}
\end{figure}

\subsection*{Cardiovascular Magnetic Resonance Imaging Reports}
We retrospectively analyzed 109 CMR imaging reports from patients scanned at the Cardiac MRI Center of the University Hospital Munster in Germany. The cohort included various cardiac diseases: hypertrophic cardiomyopathy (HCM, n=15), cardiac amyloidosis (CA, n=14), cardiac sarcoidosis (CS, n=14), myocarditis (MYO, n=14), ischemic cardiomyopathy (ICM, n=17), dilated cardiomyopathy (DCM, n=16), and normal controls (n=19). A cardiologist with extensive experience selected the reports based on well-established diagnostic criteria. All reports were written in German following the structured reporting guidelines for CMR examinations and were reviewed by a senior cardiologist. Patient consent for research use was obtained.
Each CMR report consists of two sections. The first section provides structured (primarily) quantitative data—including ventricular volumes, ejection fractions, flow measurements, quantitative mapping, and aortic area measurements. The second section is less structured narrative text describing diagnostic findings and observations.

\subsection*{Preprocessing }
The first pre-processing step includes the extraction of the relevant part of the text from the clinical report. This step involves splitting the report into the first (quantitative) and second (narrative) part. The latter section of the report is free-text describing the major findings and observations from the images and quantitative analyses. The second step includes translating the clinical text from German to English in order to enable analysis using different LLMs. The translation needs to be performed due to the fact that most of open-source LLMs have been trained on English text. The multi-lingual LLaMA3.3 model from Meta was used to translate text from German to English. The final step, analyzing the English text, involves setting up open-source LLMs locally and composing a prompt with the extracted context. Different steps are depicted in Figure \ref{fig:;method}.

\subsection*{Framework and Models}

To enable privacy-preserving deployment of LLMs in healthcare, we leveraged Ollama —an open-source framework for local LLM inference—eliminating the need to transfer sensitive patient data externally. This approach ensures secure processing of cardiac imaging reports while adhering to strict confidentiality requirements.
For further text evaluation, we compared the performance of nine different open-source LLMs: Meta’s LLaMA 3.2 (3B) and LLaMA 3.3 (70B) \cite{grattafiori2024llama}, Google’s Gemma 2 (27B)\cite{team2024gemma}, Mistral AI’s Mistral (7B)\cite{jiang2023mistral}, Deepseek’s distilled R1 (70B, 32B, 7B)\cite{deepseek2025deepseek}, and Alibaba’s Qwen2.5 (32B) \cite{qwen2024qwen} and QwQ models. These models include both reasoning LLMs, which are optimized for complex inference and decision-making (e.g., Deepseek, QwQ), and non-reasoning models, which prioritize speed and efficiency in text processing (e.g., LLaMA, Gemma, Mistral, and Qwen). Each model was assessed using a standardized clinical prompt that combines instruction text, the context of cardiac MRI findings, and the diagnostic question with different disease categories.

\subsection*{Statistics}
We conducted experiments to evaluate the performance of various models in identifying heart diagnoses from unstructured text. The LLMs are provided with context, which consists of descriptive findings from imaging reports, along with instructions to select the appropriate diagnostic category from a predefined list. To assess performance, the models generate responses in a structured JSON format, allowing for direct comparison with the ground truth diagnosis. The evaluation is based on metrics such as accuracy, precision, recall, and F1 score. Additionally, we measure and record the execution time for each model. All models were run on a single NVIDIA A40 GPU with 48 GB of Memory. We evaluated the performance of models against a board-certified cardiologist with one year of experience and expertise in analyzing CMR scans.

\bibliography{sample}

\section*{Acknowledgements}

This study was supported by a research grant from the "Peter-Lancier-Stiftung", Hamburg, Germany (project title "OPT-AI-CMR", ID BD6505565).

\section*{Author contributions statement}

S.A., first author, conceived the study, designed and conducted the experiments, and wrote the main manuscript text. V.V. contributed to the experimental design and provided domain expertise throughout the research process. M.B. analysed the results and contributed to interpretation and visualization of the data. A.Y. supervised the project and provided critical revisions and clinical guidance. All authors reviewed and approved the final manuscript.




\end{document}